\documentclass[abstract=on,notitlepage,twocolomn,superscriptaddress,nofootinbib,aps,showpac,twocolumn,prd]{revtex4}
\usepackage{amsfonts}
\usepackage{amsmath}
\usepackage{amssymb}
\usepackage{enumerate}
\usepackage{graphicx}
\usepackage{color}
\usepackage{epstopdf}
\usepackage{float}
\usepackage{dcolumn}
\usepackage{hyperref}
\usepackage{amsthm}

\begin{document}


\title{Gravitational Wave Heating}



\author{Vishnu Kakkat}\email[]{kakkav@unisa.ac.za}
\affiliation{Department of Mathematical sciences, Unisa}
\author{Nigel T. Bishop}\email[]{n.bishop@ru.ac.za}
\affiliation{Department of Mathematics, Rhodes University}
\author{Amos S. Kubeka}\email[]{kubekas@unisa.ac.za}
\affiliation{Department of Mathematical sciences, Unisa}

\date{\today}

\begin{abstract}
It was shown in previous work that when a gravitational wave (GW) passes through a viscous shell of matter the magnitude of the GW will be damped and there are astrophysical circumstances in which the damping is almost complete. The energy transfer from the GWs to the fluid will increase its temperature. We construct a model for this process and obtain an expression for the temperature distribution inside the shell in terms of spherical harmonics. Further, it is shown that this effect is astrophysically significant: a model problem is constructed for which the temperature increase is of order $10^6\,{}^\circ$K.
\end{abstract}
\maketitle

\section{Introduction}

Studies on gravitational waves (GWs) have garnered significant attention in recent years, primarily due to the regular direct detections. As GWs propagate from their source, they interact with matter in various ways, and some of these interactions can be found in~\cite{hawking1966perturbations,esposito1971absorption,marklund2000radio,brodin2001photon,cuesta2002gravitational}. Despite their interactions, GWs are usually unaffected by matter, allowing them to traverse cosmological distances without significant attenuation. It is well-known that when GWs propagate through a perfect fluid, they do not experience any absorption or dissipation, and further that when passing through a viscous fluid, energy is transferred from the GWs to the fluid~\cite{hawking1966perturbations}. It has been common practice to take the rate of energy transfer as
\begin{equation}
\frac{d\dot{E}_{GW}}{dr}=16\pi\eta \dot{E}_{GW}
\label{eqn:eta}
\end{equation}
in geometric units where $\eta$ is the viscosity and $r$ is distance. 

However, recent studies~\cite{bishop2020effect,bishop2022effect} have shown that a shell composed of viscous fluid surrounding a GW event modifies the magnitude of the GWs according to a formula that reduces to Eq.~\eqref{eqn:eta} when the matter is far from the GW source, but can be much larger when the matter is at a distance comparable to the wavelength.

Building upon this concept, we investigate the behavior of a spherically symmetric viscous shell around a circular binary source when traversed by GWs. We observed that GWs cause the shell to heat up, which may lead to the emission of electromagnetic (EM) waves. To explore the astrophysical significance of this heating mechanism, we applied it to a model of a stationary accretion disk.

Accretion disks are a common and widely observed astrophysical phenomenon, and various models for accretion disks have been proposed. A detailed review on accretion disk models can be found in~\cite{abramowicz2013foundations}. In our work, we investigated the implications of the GW heating mechanism by considering a stationary accretion disk located at a finite distance from the source of GWs. A similar prediction to ours was made in~\cite{kocsis2008brightening}, demonstrating the brightening of an accretion disk close to a binary black hole merger when the shell radius is much larger than the wavelength. This study can be seen as the limiting case of~\cite[Eq.19]{bishop2022effect}. However, the current study is more general as it allows for the variation of the viscous heating effect with distance from the GW source. We derive an expression for the temperature rise within the shell, expressed in terms of spherical harmonics. Importantly, our findings align with the expression given in~\cite[Eq. 2]{kocsis2008brightening} in the limit of being far from the source.

Previous studies have demonstrated that GW heating effects can result in an EM burst, as shown in~\cite{milosavljevic2005afterglow,tanaka2010time,li2012gravitational}. However, it remains uncertain whether this gamma-ray burst can be observed or detected. Furthermore, it is reported that the afterglow can be observable as a rapidly brightening source soon after the merger~\cite{tanaka2010time}. Similar to the case described in~\cite{kocsis2008brightening}, it has been argued in~\cite{li2012gravitational} that GW heating luminosities of the accretion disk and stars are low and may not lead to significant EM flares relative to their intrinsic luminosity, except in certain cases. Nevertheless, the GW heating effect is a significant astrophysical phenomena and needs to be considered during astrophysical observations.

This paper is organized as follows: Section (\ref{sec:previous}) outlines previous work, and in section \ref{sec:heating}, we derive expressions, under different conditions, for the temperature increases in matter around a GW source. Next, in section \ref{sec:astrophysics}, we apply the model to a specific astrophysical problem to understand the significance of the effect. Finally, in section \ref{sec:summary}, we provide a summary and conclusion of our findings. Appendix~\ref{a-cas} presents the computer scripts used to derive some of the results in the paper.

\section{Previous work}\label{sec:previous}

The Bondi-Sachs formalism is a well-known mathematical framework used in general relativity \cite{bishop1999incorporation,bishop1997high,gomez2001gravitational}. 
Consider the Bondi-Sachs metric representing a general spacetime \cite{sachs1962gravitational,bondi1962gravitational} in null coordinates $(u,r,x^A)$ as
\begin{align}
ds^2=-\left(e ^{2\beta}(1+W_{c}r)-r^2h_{AB}U^AU^B\right)du^2-2e^{2\beta}dudr\\-r^2h_{AB}U^B dudx^A+r^2h_{AB}dx^Adx^B,\nonumber
\end{align}
where $h_{AB}h^{BC}=\delta_C ^A$, $\det(h_{AB})=\det(q_{AB})$, $q_{AB}$ is the canonical metric on the unit sphere. Here coordinate $u$ labels the null outgoing hypersurface, $r$ coordinate is the surface area coordinate, and $x^A=(\theta,\phi)$ are the spherical polar coordinates.\\
Let $q_A$ be a complex dyad and defined as 
\begin{equation}
q^A=(1,\frac{i}{\sin\theta}), \qquad q_A=(1,i\sin\theta) 
\end{equation}
Then the $h_{AB}$ can be represented as 
\begin{equation}
J=h_{AB}\frac{q^Aq^B}{2}.
\end{equation}
Notice that $J=0$ represents the spherically symmetric spacetime. Let the spin-weighted field $U$ be defined by
\begin{equation}
U=U^Aq_{A}
\end{equation}
and similarly, we can define the complex differential operators $\eth,\bar{\eth}$. see \cite{gomez1997eth,bishop1999incorporation,newman1966note} for a more detailed explanation.

We make the ansatz of small quadrupolar perturbations about Minkowski spacetime with the metric quantities $\beta,U,W_c,J$ taking the form
\begin{align}
\beta=&\Re(\beta^{[2,2]}(r)e^{i\nu u}){}_0Z_{2,2}\,,\;\;
U=\Re(U^{[2,2]}(r)e^{i\nu u}){}_1Z_{2,2}\,,\nonumber \\
W_c=&\Re(W_c^{[2,2]}(r)e^{i\nu u}){}_0Z_{2,2}\,,\;\;
J=\Re(J^{[2,2]}(r)e^{i\nu u}){}_2Z_{2,2}\,.
\label{e-ansatz}
\end{align}
The perturbations oscillate in time with frequency $\nu/(2\pi)$. The quantities ${}_s Z_{\ell,m}$ are spin-weighted spherical harmonic basis functions related to the usual ${}_s Y_{\ell,m}$ as specified in~\cite{bishop2005linearized,bishop2016extraction}. They have the property that ${}_0 Z_{\ell,m}$ are real, enabling the description of the metric quantities $\beta,W_c$ (which are real) without mode-mixing; however, for $s\ne 0$ ${}_s Z_{2,2}$ is, in general, complex. A general solution may be constructed by summing over the $(\ell,m)$ modes.
As shown in previous work~\cite{bishop2005linearized,bishop2020effect}, solving the vacuum Einstein equations under the condition of no incoming radiation leads to
\begin{align}
\beta^{[2,2]} ={}& b_0, \nonumber \\
W_c^{[2,2]} ={}& 4i\nu b_0 - 2\nu^4 C_{40} - 2\nu^2 C_{30} + \frac{4i\nu C_{30} - 2b_0 }{r} \nonumber \\
&+\frac{4i\nu^3 C_{40}}{r}+ \frac{12 \nu^2 C_{40}}{r^2} - \frac{12i\nu C_{40}}{r^3} - \frac{6C_{40}}{r^4}, \nonumber \\
U^{[2,2]} ={}& \frac{\sqrt{6}(-2i\nu b_0 + \nu^4 C_{40} + \nu^2 C_{30})}{3} + \frac{2\sqrt{6} b_0}{r} \nonumber \\
&+ \frac{2\sqrt{6} C_{30}}{r^2} - \frac{4i\nu\sqrt{6} C_{40}}{r^4} - \frac{3\sqrt{6} C_{40}}{r^4}, \nonumber \\
J^{[2,2]} ={}& \frac{2\sqrt{6}(2b_0 + i\nu^3 C_{40} + i\nu C_{30})}{3} + \frac{2\sqrt{6} C_{30}}{r} \nonumber \\& + \frac{2\sqrt{6} C_{40}}{r^3},
\label{e-pert}
\end{align}
with constants of integration $b_0, C_{30}, C_{40}$. Denoting the news for the solution Eq.~(\ref{e-pert}) by ${\mathcal N}_0$, and allowing for the conventions used here, we find ${\mathcal N}_{0}=-\sqrt{6}\nu^3 \Re(iC_{40}\exp(i\nu u))\,{}_2Z_{2,2}$. Thus the constant $C_{40}$ is physical and represents the magnitude of the GW source, while the constants $b_0$ and $C_{30}$ represent gauge freedoms.

We now consider the case that the GW source is surrounded by a shell of matter. Due to the GW perturbations, the matter within the shell undergoes motion, and the velocity field is calculated using the matter conservation conditions~\cite{bishop2022effect}. Having found the velocity field, it is then straightforward to calculate the shear tensor $\sigma_{ab}$, and it was shown~\cite{bishop2022effect} that
\begin{align}\label{sigmacal}
\sigma_{00} &= \sigma_{01} = \sigma_{0A} = 0\,, \nonumber \\
-\sigma_{11}^{[2,2]} &= h^{AB}\sigma_{AB}^{[2,2]} \nonumber \\
&= \Re\Bigr(12 C_{40}\frac{3i-3r\nu-ir^2\nu^2}{\nu r^5}\exp(i\nu u)\Bigl)Z_{2,2}\,, \nonumber \\
q^A \sigma_{1A}^{[2,2]} &= \Re\Bigr(2 C_{40}\frac{6i-6\nu r-3i\nu^2r^2+\nu^3r^3}{\nu r^4}\exp(i\nu u)\Bigl){}_1Z_{2,2}\,, \nonumber \\
q^A q^B \sigma_{AB}^{[2,2]} &= \Re\Bigr(C_{40}\frac{-3-3i\nu r+3\nu^2r^2+2i\nu^3r^3-\nu^4r^4}{\nu r^4}\nonumber \\
&\quad\exp(i\nu u)\Bigl)\times{}_2Z_{2,2}. 
\end{align}
It is shown in \cite{baumgarte2010numerical} that
\begin{equation}\label{eqn:Et}
\frac{\partial_u E_{shell}}{\Delta V}=2\eta\sigma_{ab}\sigma^{ab}\,,
\end{equation}
where $t=u+r$, $E_{shell}$ is the energy in an element of the shell with volume $\Delta V$, and $\eta$ is the coefficient of (dynamic) viscosity.

\section{The heating effect}\label{sec:heating}
We now investigate the GW heating effect on a shell of matter surrounding a source that comprises a circular binary. As was shown previously~\cite{bishop2011initial}, the perturbative quantities of Eq.~(\ref{e-ansatz}) are amended to
\begin{equation}
J=\Re(J^{[2,2]}(r)e^{i\nu u}){}_2Z_{2,2}-\Re(iJ^{[2,2]}(r)e^{i\nu u}){}_2Z_{2,-2}\,,
\end{equation}
with similar expressions for $\beta,U,W_c$, and also for the shear expressions of Eq.~(\ref{sigmacal}). In all cases, the coefficients of ${}_sZ_{2,2}$ and ${}_sZ_{2,-2}$ have the same $r$-behaviour but are out of phase in time $u$.

The computer algebra evaluates $\sigma_{ab}\sigma^{ab}$ in Eq.~(\ref{eqn:Et}), and the resulting expression is lengthy. However, it is greatly simplified if time-averaging is applied, i.e. we evaluate
\begin{equation}\label{eqn:EtTA}
\left<\frac{\partial_u E_{shell}}{\Delta V}\right>=<2\eta\sigma_{ab}\sigma^{ab}>=\frac{\nu}{2\pi}\int_0^{2\pi/\nu}2\eta\sigma_{ab}\sigma^{ab}du\,;
\end{equation}
note that time-averaging means that the results to be obtained apply only on a time-scale that is greater than the averaging period of $2\pi/\nu$. We find
\begin{multline}\label{souceenergy}
\left<\frac{\partial_u E_{shell}}{\Delta V}\right>=\frac{15 C_{40}^2\eta}{8\pi{\nu}^2r^{10}}\times \\ \Bigl[\bigl({\nu}^8r^8-18{\nu}^6r^6+159{\nu}^4r^4+315{\nu}^2r^2+405\bigr)\cos^4\theta\\
+\bigl(6{\nu}^8r^8-12{\nu}^6r^6-198{\nu}^4r^4-702{\nu}^2r^2-1890\bigr)\cos^2\theta\\
+{\nu}^8r^8+14{\nu}^6r^6+63{\nu}^4r^4+315{\nu}^2r^2+1557\Bigr],
\end{multline}
which is then decomposed into axisymmetric spherical harmonics $Y_{\ell,0}$
\begin{equation*}
\left<\frac{\partial_u E_{shell}}{\Delta V}\right>=
\eta C_{40}^2\nu^8\left(D_0Y_{0,0}+D_2Y_{2,0}+D_4Y_{4,0}\right)\,,
\end{equation*}
\begin{align}
D_0&= \frac{12({\nu}^8r^8+2{\nu}^6r^6+9{\nu}^4r^4+45{\nu}^2r^2+315)}{\sqrt{\pi}{\nu}^{10}r^{10}},\nonumber \\
D_2&=\frac{24 \sqrt{5}({\nu}^8r^8-4{\nu}^6r^6-9{\nu}^4r^4-63{\nu}^2r^2-225)}{7\sqrt{\pi}{\nu}^{10}r^{10}},\nonumber \\
D_4 &=\frac{2({\nu}^8r^8-18{\nu}^6r^6+159{\nu}^4r^4+315{\nu}^2r^2+405)}{7\sqrt{\pi}{\nu}^{10}r^{10}}.
\label{e-TA-D}
\end{align}

Previous work established a relation between $C_{40}$ and the rate of energy emission as GWs $\partial_u E_{GW}=3\nu^2C_{40}^2/(2\pi)$. That expression was for the case that the GW comprises a ${}_2Z_{2,2}$ component only, and here there is also a ${}_2Z_{2,-2}$ component, so we have
\begin{equation}\label{eqn:c40}
C_{40} ^2=\frac{\pi}{3\nu^6}\partial_u E_{GW}.
\end{equation}
Combining Eqs. \eqref{e-TA-D} and \eqref{eqn:c40} gives
\begin{equation}\label{eqn:eshell}
\frac{\partial_u E_{shell}}{\Delta V}=\frac{\pi}{3}\nu^2\eta \partial_u E_{GW}\left(D_0Y_{0,0}+D_2Y_{2,0}+D_4Y_{4,0}\right)\\,.
\end{equation}
The above expression \eqref{eqn:eshell} can be re-written as 
\begin{equation}\label{eqn:Eshell}
\partial_u E_{shell}=\frac{\pi\eta}{3\rho}\nu^2 \Delta m \partial_u E_{GW}\left(D_0Y_{0,0}+D_2Y_{2,0}+D_4Y_{4,0}\right)\,,
\end{equation}
where $\Delta m$ is the mass of a fluid element and $\rho$ denotes
its density.
We now need to convert Eq. \eqref{eqn:Eshell} to SI units, which means that it must be multiplied by powers of $G$ and $c$ so that $\Delta m\nu^2\eta/\rho$ become dimensionless. Therefore in SI units, Eq. \eqref{eqn:Eshell} becomes
\begin{equation}
\partial_u E_{shell}=\frac{\pi G\eta}{3 c^5\rho}\nu^2 \Delta m \partial_u E_{GW}\left(D_0Y_{0,0}+D_2Y_{2,0}+D_4Y_{4,0}\right)\,,
\end{equation}
and in the formulas for $D_i$ in Eq.~\eqref{e-TA-D}, $r\nu\rightarrow r \nu/c^2$.

Next, we note that $\partial _u E_{shell}=\Delta mC\partial_u T,$ where $C$ is the specific heat capacity and $T$ is the temperature at an event in the shell, so that
\begin{equation}
\partial_u T=\frac{\pi G\eta}{3c^5 C\rho}\nu^2  \partial_u E_{GW}\left(D_0Y_{0,0}+D_2Y_{2,0}+D_4Y_{4,0}\right)\,.
\label{e-du_T}
\end{equation}
We proceed further by considering two different cases (A) heat flow within the shell and constant GW frequency, and (B) GWs with variable frequency and no heat flow within the shell. Actually, both effects can be included and a solution obtained that can be written as a sum of integrals, but doing so makes the formulas less transparent.

\subsection{Heat flow within the shell}
Allowing for heat flow within the shell gives 
\begin{align}
\partial_u T &= \alpha\nabla ^2 T + \frac{\pi G\eta}{3c^5 C\rho}\nu^2  \partial_u E_{GW} \nonumber \\
&\quad \times \left(D_0Y_{0,0} + D_2Y_{2,0} + D_4Y_{4,0}\right).
\end{align}
where $\alpha$ is the thermal diffusivity of the matter in the shell. Then assuming $T=T_0$ at $t=0$ and using the abbreviation 
\begin{equation}
\label{eqn:A}
A=B\nu^2  \partial_u E_{GW}\,,\;\;\mbox{with }B=\frac{\pi G\eta}{3c^5 C\rho}\,,
\end{equation}
we obtain 
\begin{align}
\label{eqn:Temp}
T &= T_0 + uAD_0 Y_{0,0} + \frac{AD_2 Y_{2,0} r^2}{6\alpha}\biggl(1-e^{-6\alpha u/r^2}\biggr) \nonumber \\
&\quad + \frac{AD_4 Y_{4,0}r^2}{20\alpha}\biggl(1-e^{-20\alpha u/r^2}\biggr).
\end{align}
Eq. \eqref{eqn:Temp} above represents the temperature distribution inside the shell expressed in terms of spherical harmonics. The effect is driven by the flow of GWs through the shell, $\partial_u E_{GW}$; and the form of the temperature distribution is determined by the wave frequency $\nu$, as well as by the physical properties of the viscous shell, specifically the specific heat $C$, the thermal diffusivity $\alpha$, the viscosity $\eta$ and the density $\rho$.

It is instructive to consider two special cases of Eq.~\eqref{eqn:Temp}. Define $\mu=r^2/(\alpha u)$, and consider $\mu\gg 1$ and $\mu\ll 1$, corresponding to low and high thermal diffusivity respectively. For $\mu\gg 1$,
\begin{equation}
T=T_0+uA(D_0 Y_{0,0}+ D_2 Y_{2,0}+D_4 Y_{4,0})\,,
\label{e-dT_low_dif}
\end{equation}
and for $\mu\ll 1$
\begin{equation}
T=T_0+uA\left(D_0 Y_{0,0}+ \mu(D_2 Y_{2,0}+D_4 Y_{4,0})\right)\,.
\label{e-dT_high_dif}
\end{equation}
In the case of high thermal diffusivity, the temperature variation over the sphere is small, and in all cases the order of magnitude of the temperature change is
\begin{equation}\label{eqn:diffT}
(T-T_0)\sim uAD_0\,.
\end{equation}

In the formulas above the frequency $\nu$ is treated as constant, and the temperature change is very sensitive to the value of $r\nu$. If the frequency varies, and in order to avoid overestimating/underestimating the effect, $\nu$ should be chosen towards the top/bottom respectively of the frequency range.

\subsection{Variable frequency}
In this case, the quantities $D_0,D_2,D_4$ depend on $\nu$ and so also vary. The solution to Eq. \eqref{e-du_T} can be written
\begin{align}
&T(u_2)=T(u_1)+B\int^{u_2}_{u_1}
\nu^2(u)  \partial_u E_{GW}(u)\times\nonumber \\
&\left(D_0(u)Y_{0,0}+D_2(u)Y_{2,0}(\theta)+D_4(u)Y_{4,0}(\theta)\right) du\,,
\label{e-du_T_vf}
\end{align}
where $B$ was defined in Eq.~\eqref{eqn:A} and, in general, the integral would need to be evaluated numerically.

\subsection{Equal mass circular binary}
In the case of two equal mass binaries, we have
\begin{equation}
\partial_u E_{GW}=\frac{2M^2r_0 ^4\nu^6}{5}\,,
\end{equation}
where $r_0$ is the orbital radius (see, e.g. \cite{bishop2011initial}, but note that the formulas appear to be different because the reference uses $\nu$ as the orbital frequency rather than the wave frequency). Hence Eq. \eqref{eqn:A} becomes
\begin{equation}
A=\frac{8\pi GM^2 r_0 ^4\nu^8\eta}{15c^5 C\rho}\,,
\end{equation}
and this form of $A$ is used in Eq. \eqref{eqn:Temp} to determine the temperature distribution in the shell.

\section{Relevance to astrophysics}\label{sec:astrophysics}

A key question is whether there are astrophysical circumstances such that the temperature increase would be large enough to be significant. Here, we describe one scenario in which that would be the case, so motivating the astrophysical importance of the GW heating effect. We consider the merger of two black holes, and note the observed parameters from GW150914 \cite{scientific2016tests}
\begin{align}
\Delta E_{GW}&=3M_\odot=5.36\times 10^{47} \mbox{J}\,,
f=132 \mbox{Hz}\,,\nonumber \\
M_f&=62 M_\odot\,,\left(\partial_u E_{GW}\right)_{\mbox{peak}}=200 M_\odot/\mbox{s}
\end{align}
where $f$ is the frequency at merger and in the formulas above $\nu=2\pi f$; and $M_f$ is the final mass. The energy loss $\Delta E_{GW}$ is for the whole inspiral. Using a waveform from a best-fit model~\cite{GW150914}, we find that $2M_\odot$ was radiated away during the $17.6$ms between $u=0.4069$ and $u=0.4245$; during this period, the frequency increased from 90Hz through peak emission at 132Hz and increased towards 220Hz as merger gave way to ringdown. The heating effect was estimated using the variable frequency expression~\ref{e-du_T_vf}; note that the use of \eqref{e-dT_low_dif} with a fixed frequency of 155Hz (i.e., in the middle of the frequency range) led to very similar results. 

It is further supposed that matter is present in the system, and we use parameters of a stationary accretion model, as outlined in ~\cite{shakura1976theory,arai1995accretion,abramowicz2013foundations}: at the ISCO (Innermost Stable Circular Orbit), the dynamical viscosity $\eta$ is approximated as $3.5\times 10^9$J sec/m${}^3$, the density as $\rho$ as $4$kg/m${}^3$ and the specific heat as $1.43\times 10^4$J/kg/${}^\circ$K. The radius of the ISCO is taken as $r=549$km, being the value for a Schwarzschild black hole of mass $M_f$.

\begin{figure}
\includegraphics[width=8cm]{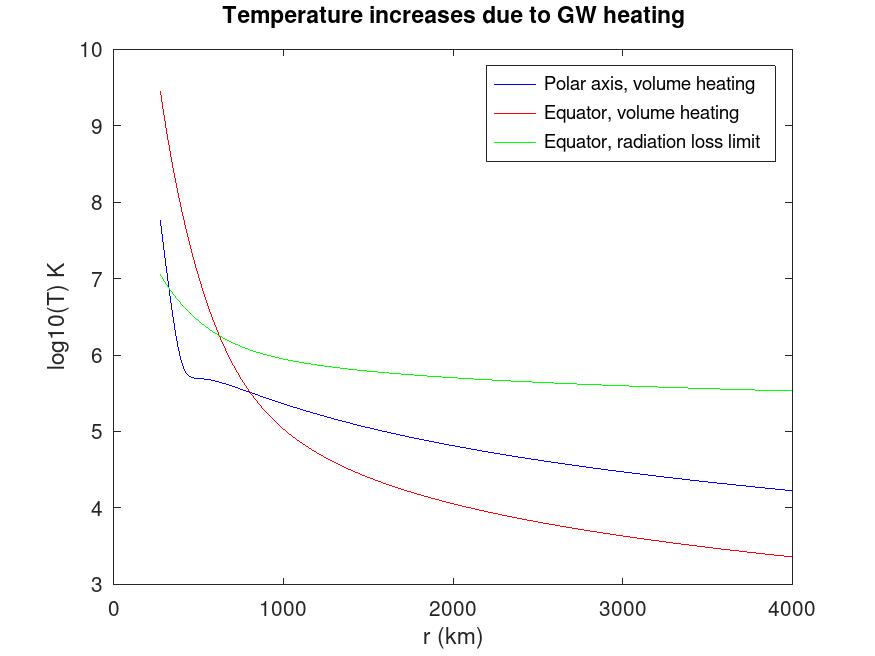}
\caption{The temperature increase (${}^\circ$K on a $\log_{10}$ scale) is plotted agains radius (km) for the cases: GW heating without radiation loss for matter on the orbital axis (blue curve); GW heating without radiation loss for matter in the equatorial plane (red curve); limitation, due to radiation loss, on temperature increase in the equatorial plane (green curve).}
\label{f-TI}
\end{figure}

The magnitude of the heating effect was evaluated for values of $r$ in the range $275$km (i.e., half of $r_{\mbox{\tiny ISCO}}$) to $3000$km, and for $\theta=0$ (on the polar axis) and $\theta=\pi/2$  (in the equatorial plane).
The GW heating effect is very sensitive to the value of $r$, and also depends on $\theta$. While we would expect an accretion disk to have $\theta\approx\pi/2$, we also evaluate the effect for matter on the polar axes ($\theta=0,\pi$). Results are shown in Fig.~\ref{f-TI}. It is noteworthy that for small $r$ the effect is much larger on the equator than at $\theta=0$. However, the situation is reversed for larger $r$, and we have checked that as $r\rightarrow\infty$ the heating effect at $\theta=0$ is 8 times that at $\theta=\pi/2$, as expected for the angular distribution of GW power of an orbiting binary. 

Energy may be radiated away, so limiting the temperature increase. Modelling the accretion disk as a disk of thickness $2h$ with $h\approx 100$km, and denoting Stefan's constant as $\sigma=5.67\times 10^{-8}$W/m${}^2$/${}^\circ$K${}^4$, it follows that the temperature increase would be limited to
\begin{equation}
(T-T_0)_{max}=\sqrt[4]{\frac{h \Delta E_{shell}}{\sigma\Delta u}}\,,
\end{equation}
where $\Delta E_{shell}$ is the energy input to a volume element of the shell in the time period $\Delta u$. The graph of $(T-T_0)_{max}$ is included in Fig.~\ref{f-TI} for the case $\theta=\pi/2$ (i.e., the equatorial plane). For $r\gtrapprox r_{\mbox{\tiny ISCO}}$, radiation loss does not limit the temperature increase due to volume heating, but it does do so for $r\lessapprox r_{\mbox{\tiny ISCO}}$. Thus the temperature increase for matter at $r=r_{\mbox{\tiny ISCO}}$ is limited to $\mathcal{O}(10^6){}^\circ$K, although matter at $r=r_{\mbox{\tiny ISCO}}/2$ could reach $10^7{}^\circ$K: there may be X-ray emission but not a gamma-ray burst.

It should also be noted that the temperature increase $T-T_0$ depends linearly on $\eta/(\rho C)$; and that it is inversely proportional to $M_f$, so that the effect would be nearly 4 times larger at the lower limit of observed black hole mergers ($M_f\approx16.7M_\odot$), and would be much smaller for supermassive black hole mergers.

In an actual black hole merger, it is expected that the inspiral of the black holes would clear out any matter in their vicinity, and there has been no astrophysical evidence of the effects of matter in an observed merger (apart, perhaps, from the Fermi observation coincident with GW150914~\cite{connaughton2016fermi}). Thus, the presence of matter around a black hole merger is highly unlikely. Further, even if matter is present, it is known that accretion disks have temperatures of the order of $10^6\,{}^\circ$K. Thus, if EM emissions are observed at a GW event corresponding to a black hole merger, it would be difficult to determine whether or not it was (partially) caused by GWs. Our purpose in presenting Fig.~\ref{f-TI} is to demonstrate that GW heating may be astrophysically significant.

\section{Summary and conclusions}\label{sec:summary}

In this article, we have derived formulas for temperature increases within a shell of viscous matter through which GWs propagate. The temperature distribution is expressed using axisymmetric spherical harmonics $Y_{\ell,0}$, with $\ell=0,\,2$ and 4; and depends on physical parameters including the viscosity $\eta$, specific heat capacity $C$, thermal diffusivity $\alpha$, and density $\rho$.

First, we considered the case of constant frequency and non-zero thermal diffusivity so that there is heat flow within the shell, and obtained Eq.~\eqref{eqn:Temp}. Simple approximations to this result were presented for the cases of low and high thermal diffusivity, Eqs.~\eqref{e-dT_low_dif} and \eqref{e-dT_high_dif} respectively. In both cases, the order of magnitude of the temperature change effect is given by Eq.~\eqref{eqn:diffT}.

We next considered the case that the GW frequency varies with time, but took the thermal diffusivity as negligible. This case is astrophysically important, since it applies to GW events caused by an inspiral and merger. The resulting temperature increase is expressed as a time-integral, Eq.~\eqref{e-du_T_vf}.

To understand the physical implications of the temperature rise, we considered the stationary accretion disk problem in a model that uses data from the binary black hole merger GW 150914. We found that the temperature rise inside the disk can be significant, being of order $\mathcal{O}(10^6){}^\circ$K. This result highlights the importance of considering this effect in astrophysical phenomena and cosmology, and in particular that previous results on GW heating in accretion disks should be revisited using formulas that properly allow for variation of the effect with distance from the source.

Additionally, we envision that GW heating may be relevant to core-collapse supernovae as well as to primordial gravitational waves~\cite{bishop2022effect}. However, the application of GW heating to various astrophysical and cosmological scenarios is beyond the scope of this paper and will be further addressed in forthcoming work.

\vspace{0.5cm}
\begin{acknowledgments}
VK and ASK express their sincere gratitude to Unisa for the Postdoctoral grant and their generous support. NTB thanks Unisa, Inter-University Centre for Astronomy and Astrophysics, and International Centre for Theoretical Sciences for hospitality.
\end{acknowledgments}

\appendix
\section{Computer scripts}
\label{a-cas}
The computer scripts are written in plain text format, and are available as Supplementary Material. Eqs.~\eqref{souceenergy} and \eqref{e-TA-D} were derived using the computer algebra system MAPLE. The file driving the calculation is \texttt{GW\_Heating.map}, which takes input from \texttt{gamma.out, initialize.map, lin.map} and \texttt{ProcRules.map}. The scripts are adapted from those reported in previous work~\cite{bishop2022effect}. The output is in \texttt{GW\_Heating.out}, and may be viewed using a plain text editor with line-wrapping switched off.

The MATLAB/Octave script \texttt{TempInc.m} performs the calculations used to produce Fig.~\ref{f-TI}.

\bibliography{vis.bib}

\begin{thebibliography}{28}
\expandafter\ifx\csname natexlab\endcsname\relax\def\natexlab#1{#1}\fi
\expandafter\ifx\csname bibnamefont\endcsname\relax
  \def\bibnamefont#1{#1}\fi
\expandafter\ifx\csname bibfnamefont\endcsname\relax
  \def\bibfnamefont#1{#1}\fi
\expandafter\ifx\csname citenamefont\endcsname\relax
  \def\citenamefont#1{#1}\fi
\expandafter\ifx\csname url\endcsname\relax
  \def\url#1{\texttt{#1}}\fi
\expandafter\ifx\csname urlprefix\endcsname\relax\def\urlprefix{URL }\fi
\providecommand{\bibinfo}[2]{#2}
\providecommand{\eprint}[2][]{\url{#2}}

\bibitem[{\citenamefont{Hawking}(1966)}]{hawking1966perturbations}
\bibinfo{author}{\bibfnamefont{S.~W.} \bibnamefont{Hawking}},
  \bibinfo{journal}{Astrophysical Journal, vol. 145, p. 544}
  \textbf{\bibinfo{volume}{145}}, \bibinfo{pages}{544} (\bibinfo{year}{1966}).

\bibitem[{\citenamefont{Esposito}(1971)}]{esposito1971absorption}
\bibinfo{author}{\bibfnamefont{F.~P.} \bibnamefont{Esposito}},
  \bibinfo{journal}{Astrophysical Journal, vol. 165, p. 165}
  \textbf{\bibinfo{volume}{165}}, \bibinfo{pages}{165} (\bibinfo{year}{1971}).

\bibitem[{\citenamefont{Marklund et~al.}(2000)\citenamefont{Marklund, Brodin,
  and Dunsby}}]{marklund2000radio}
\bibinfo{author}{\bibfnamefont{M.}~\bibnamefont{Marklund}},
  \bibinfo{author}{\bibfnamefont{G.}~\bibnamefont{Brodin}}, \bibnamefont{and}
  \bibinfo{author}{\bibfnamefont{P.~K.} \bibnamefont{Dunsby}},
  \bibinfo{journal}{The Astrophysical Journal} \textbf{\bibinfo{volume}{536}},
  \bibinfo{pages}{875} (\bibinfo{year}{2000}).

\bibitem[{\citenamefont{Brodin et~al.}(2001)\citenamefont{Brodin, Marklund, and
  Servin}}]{brodin2001photon}
\bibinfo{author}{\bibfnamefont{G.}~\bibnamefont{Brodin}},
  \bibinfo{author}{\bibfnamefont{M.}~\bibnamefont{Marklund}}, \bibnamefont{and}
  \bibinfo{author}{\bibfnamefont{M.}~\bibnamefont{Servin}},
  \bibinfo{journal}{Physical Review D} \textbf{\bibinfo{volume}{63}},
  \bibinfo{pages}{124003} (\bibinfo{year}{2001}).

\bibitem[{\citenamefont{Cuesta}(2002)}]{cuesta2002gravitational}
\bibinfo{author}{\bibfnamefont{H.~J.~M.} \bibnamefont{Cuesta}},
  \bibinfo{journal}{Physical Review D} \textbf{\bibinfo{volume}{65}},
  \bibinfo{pages}{064009} (\bibinfo{year}{2002}).

\bibitem[{\citenamefont{Bishop et~al.}(2020)\citenamefont{Bishop, van~der Walt,
  and Naidoo}}]{bishop2020effect}
\bibinfo{author}{\bibfnamefont{N.~T.} \bibnamefont{Bishop}},
  \bibinfo{author}{\bibfnamefont{P.~J.} \bibnamefont{van~der Walt}},
  \bibnamefont{and} \bibinfo{author}{\bibfnamefont{M.}~\bibnamefont{Naidoo}},
  \bibinfo{journal}{General Relativity and Gravitation}
  \textbf{\bibinfo{volume}{52}}, \bibinfo{pages}{92} (\bibinfo{year}{2020}).

\bibitem[{\citenamefont{Bishop et~al.}(2022)\citenamefont{Bishop, van~der Walt,
  and Naidoo}}]{bishop2022effect}
\bibinfo{author}{\bibfnamefont{N.~T.} \bibnamefont{Bishop}},
  \bibinfo{author}{\bibfnamefont{P.~J.} \bibnamefont{van~der Walt}},
  \bibnamefont{and} \bibinfo{author}{\bibfnamefont{M.}~\bibnamefont{Naidoo}},
  \bibinfo{journal}{Physical Review D} \textbf{\bibinfo{volume}{106}},
  \bibinfo{pages}{084018} (\bibinfo{year}{2022}).

\bibitem[{\citenamefont{Abramowicz and
  Fragile}(2013)}]{abramowicz2013foundations}
\bibinfo{author}{\bibfnamefont{M.~A.} \bibnamefont{Abramowicz}}
  \bibnamefont{and} \bibinfo{author}{\bibfnamefont{P.~C.}
  \bibnamefont{Fragile}}, \bibinfo{journal}{Living Reviews in Relativity}
  \textbf{\bibinfo{volume}{16}}, \bibinfo{pages}{1} (\bibinfo{year}{2013}).

\bibitem[{\citenamefont{Kocsis and Loeb}(2008)}]{kocsis2008brightening}
\bibinfo{author}{\bibfnamefont{B.}~\bibnamefont{Kocsis}} \bibnamefont{and}
  \bibinfo{author}{\bibfnamefont{A.}~\bibnamefont{Loeb}},
  \bibinfo{journal}{Physical Review Letters} \textbf{\bibinfo{volume}{101}},
  \bibinfo{pages}{041101} (\bibinfo{year}{2008}).

\bibitem[{\citenamefont{Milosavljevi{\'c} and
  Phinney}(2005)}]{milosavljevic2005afterglow}
\bibinfo{author}{\bibfnamefont{M.}~\bibnamefont{Milosavljevi{\'c}}}
  \bibnamefont{and} \bibinfo{author}{\bibfnamefont{E.~S.}
  \bibnamefont{Phinney}}, \bibinfo{journal}{The Astrophysical Journal}
  \textbf{\bibinfo{volume}{622}}, \bibinfo{pages}{L93} (\bibinfo{year}{2005}).

\bibitem[{\citenamefont{Tanaka and Menou}(2010)}]{tanaka2010time}
\bibinfo{author}{\bibfnamefont{T.}~\bibnamefont{Tanaka}} \bibnamefont{and}
  \bibinfo{author}{\bibfnamefont{K.}~\bibnamefont{Menou}},
  \bibinfo{journal}{The Astrophysical Journal} \textbf{\bibinfo{volume}{714}},
  \bibinfo{pages}{404} (\bibinfo{year}{2010}).

\bibitem[{\citenamefont{Li et~al.}(2012)\citenamefont{Li, Kocsis, and
  Loeb}}]{li2012gravitational}
\bibinfo{author}{\bibfnamefont{G.}~\bibnamefont{Li}},
  \bibinfo{author}{\bibfnamefont{B.}~\bibnamefont{Kocsis}}, \bibnamefont{and}
  \bibinfo{author}{\bibfnamefont{A.}~\bibnamefont{Loeb}},
  \bibinfo{journal}{Monthly Notices of the Royal Astronomical Society}
  \textbf{\bibinfo{volume}{425}}, \bibinfo{pages}{2407} (\bibinfo{year}{2012}).

\bibitem[{\citenamefont{Bishop et~al.}(1999)\citenamefont{Bishop, G{\'o}mez,
  Lehner, Maharaj, and Winicour}}]{bishop1999incorporation}
\bibinfo{author}{\bibfnamefont{N.~T.} \bibnamefont{Bishop}},
  \bibinfo{author}{\bibfnamefont{R.}~\bibnamefont{G{\'o}mez}},
  \bibinfo{author}{\bibfnamefont{L.}~\bibnamefont{Lehner}},
  \bibinfo{author}{\bibfnamefont{M.}~\bibnamefont{Maharaj}}, \bibnamefont{and}
  \bibinfo{author}{\bibfnamefont{J.}~\bibnamefont{Winicour}},
  \bibinfo{journal}{Physical Review D} \textbf{\bibinfo{volume}{60}},
  \bibinfo{pages}{024005} (\bibinfo{year}{1999}).

\bibitem[{\citenamefont{Bishop et~al.}(1997)\citenamefont{Bishop, G{\'o}mez,
  Lehner, Maharaj, and Winicour}}]{bishop1997high}
\bibinfo{author}{\bibfnamefont{N.~T.} \bibnamefont{Bishop}},
  \bibinfo{author}{\bibfnamefont{R.}~\bibnamefont{G{\'o}mez}},
  \bibinfo{author}{\bibfnamefont{L.}~\bibnamefont{Lehner}},
  \bibinfo{author}{\bibfnamefont{M.}~\bibnamefont{Maharaj}}, \bibnamefont{and}
  \bibinfo{author}{\bibfnamefont{J.}~\bibnamefont{Winicour}},
  \bibinfo{journal}{Physical Review D} \textbf{\bibinfo{volume}{56}},
  \bibinfo{pages}{6298} (\bibinfo{year}{1997}).

\bibitem[{\citenamefont{G{\'o}mez}(2001)}]{gomez2001gravitational}
\bibinfo{author}{\bibfnamefont{R.}~\bibnamefont{G{\'o}mez}},
  \bibinfo{journal}{Physical Review D} \textbf{\bibinfo{volume}{64}},
  \bibinfo{pages}{024007} (\bibinfo{year}{2001}).

\bibitem[{\citenamefont{Sachs}(1962)}]{sachs1962gravitational}
\bibinfo{author}{\bibfnamefont{R.~K.} \bibnamefont{Sachs}},
  \bibinfo{journal}{Proceedings of the Royal Society of London. Series A.
  Mathematical and Physical Sciences} \textbf{\bibinfo{volume}{270}},
  \bibinfo{pages}{103} (\bibinfo{year}{1962}).

\bibitem[{\citenamefont{Bondi et~al.}(1962)\citenamefont{Bondi, Van~der Burg,
  and Metzner}}]{bondi1962gravitational}
\bibinfo{author}{\bibfnamefont{H.}~\bibnamefont{Bondi}},
  \bibinfo{author}{\bibfnamefont{M.~G.~J.} \bibnamefont{Van~der Burg}},
  \bibnamefont{and} \bibinfo{author}{\bibfnamefont{A.}~\bibnamefont{Metzner}},
  \bibinfo{journal}{Proceedings of the Royal Society of London. Series A.
  Mathematical and Physical Sciences} \textbf{\bibinfo{volume}{269}},
  \bibinfo{pages}{21} (\bibinfo{year}{1962}).

\bibitem[{\citenamefont{G{\'o}mez et~al.}(1997)\citenamefont{G{\'o}mez, Lehner,
  Papadopoulos, and Winicour}}]{gomez1997eth}
\bibinfo{author}{\bibfnamefont{R.}~\bibnamefont{G{\'o}mez}},
  \bibinfo{author}{\bibfnamefont{L.}~\bibnamefont{Lehner}},
  \bibinfo{author}{\bibfnamefont{P.}~\bibnamefont{Papadopoulos}},
  \bibnamefont{and} \bibinfo{author}{\bibfnamefont{J.}~\bibnamefont{Winicour}},
  \bibinfo{journal}{Classical and Quantum Gravity}
  \textbf{\bibinfo{volume}{14}}, \bibinfo{pages}{977} (\bibinfo{year}{1997}).

\bibitem[{\citenamefont{Newman and Penrose}(1966)}]{newman1966note}
\bibinfo{author}{\bibfnamefont{E.~T.} \bibnamefont{Newman}} \bibnamefont{and}
  \bibinfo{author}{\bibfnamefont{R.}~\bibnamefont{Penrose}},
  \bibinfo{journal}{Journal of Mathematical Physics}
  \textbf{\bibinfo{volume}{7}}, \bibinfo{pages}{863} (\bibinfo{year}{1966}).

\bibitem[{\citenamefont{Bishop}(2005)}]{bishop2005linearized}
\bibinfo{author}{\bibfnamefont{N.~T.} \bibnamefont{Bishop}},
  \bibinfo{journal}{Classical and Quantum Gravity}
  \textbf{\bibinfo{volume}{22}}, \bibinfo{pages}{2393} (\bibinfo{year}{2005}).

\bibitem[{\citenamefont{Bishop and Rezzolla}(2016)}]{bishop2016extraction}
\bibinfo{author}{\bibfnamefont{N.~T.} \bibnamefont{Bishop}} \bibnamefont{and}
  \bibinfo{author}{\bibfnamefont{L.}~\bibnamefont{Rezzolla}},
  \bibinfo{journal}{Living reviews in relativity}
  \textbf{\bibinfo{volume}{19}}, \bibinfo{pages}{1} (\bibinfo{year}{2016}).

\bibitem[{\citenamefont{Baumgarte and Shapiro}(2010)}]{baumgarte2010numerical}
\bibinfo{author}{\bibfnamefont{T.~W.} \bibnamefont{Baumgarte}}
  \bibnamefont{and} \bibinfo{author}{\bibfnamefont{S.~L.}
  \bibnamefont{Shapiro}}, \emph{\bibinfo{title}{Numerical relativity: solving
  Einstein's equations on the computer}} (\bibinfo{publisher}{Cambridge
  University Press}, \bibinfo{year}{2010}).

\bibitem[{\citenamefont{Bishop et~al.}(2011)\citenamefont{Bishop, Pollney, and
  Reisswig}}]{bishop2011initial}
\bibinfo{author}{\bibfnamefont{N.}~\bibnamefont{Bishop}},
  \bibinfo{author}{\bibfnamefont{D.}~\bibnamefont{Pollney}}, \bibnamefont{and}
  \bibinfo{author}{\bibfnamefont{C.}~\bibnamefont{Reisswig}},
  \bibinfo{journal}{Classical and Quantum Gravity}
  \textbf{\bibinfo{volume}{28}}, \bibinfo{pages}{155019}
  (\bibinfo{year}{2011}).

\bibitem[{\citenamefont{Scientific et~al.}(2016)\citenamefont{Scientific,
  Collaborations, Abbott, Abbott, Abbott, Abernathy, Acernese, Ackley, Adams,
  Adams et~al.}}]{scientific2016tests}
\bibinfo{author}{\bibfnamefont{L.}~\bibnamefont{Scientific}},
  \bibinfo{author}{\bibfnamefont{V.}~\bibnamefont{Collaborations}},
  \bibinfo{author}{\bibfnamefont{B.}~\bibnamefont{Abbott}},
  \bibinfo{author}{\bibfnamefont{R.}~\bibnamefont{Abbott}},
  \bibinfo{author}{\bibfnamefont{T.}~\bibnamefont{Abbott}},
  \bibinfo{author}{\bibfnamefont{M.}~\bibnamefont{Abernathy}},
  \bibinfo{author}{\bibfnamefont{F.}~\bibnamefont{Acernese}},
  \bibinfo{author}{\bibfnamefont{K.}~\bibnamefont{Ackley}},
  \bibinfo{author}{\bibfnamefont{C.}~\bibnamefont{Adams}},
  \bibinfo{author}{\bibfnamefont{T.}~\bibnamefont{Adams}},
  \bibnamefont{et~al.}, \bibinfo{journal}{Physical review letters}
  \textbf{\bibinfo{volume}{116}}, \bibinfo{pages}{221101}
  (\bibinfo{year}{2016}).

\bibitem[{\citenamefont{LSC}(2020)}]{GW150914}
\bibinfo{author}{\bibnamefont{LSC}}, \emph{\bibinfo{title}{Gravitational wave
  open science center}} (\bibinfo{year}{2020}), \bibinfo{note}{accessed on 3
  December 2023},
  \urlprefix\url{https://gwosc.org/s/events/GW150914/P150914/fig2-unfiltered-waveform-H.txt}.

\bibitem[{\citenamefont{Shakura and Sunyaev}(1976)}]{shakura1976theory}
\bibinfo{author}{\bibfnamefont{N.}~\bibnamefont{Shakura}} \bibnamefont{and}
  \bibinfo{author}{\bibfnamefont{R.}~\bibnamefont{Sunyaev}},
  \bibinfo{journal}{Monthly Notices of the Royal Astronomical Society}
  \textbf{\bibinfo{volume}{175}}, \bibinfo{pages}{613} (\bibinfo{year}{1976}).

\bibitem[{\citenamefont{Arai and Hashimoto}(1995)}]{arai1995accretion}
\bibinfo{author}{\bibfnamefont{K.}~\bibnamefont{Arai}} \bibnamefont{and}
  \bibinfo{author}{\bibfnamefont{M.}~\bibnamefont{Hashimoto}},
  \bibinfo{journal}{Astronomy and Astrophysics, v. 302, p. 99}
  \textbf{\bibinfo{volume}{302}}, \bibinfo{pages}{99} (\bibinfo{year}{1995}).

\bibitem[{\citenamefont{Connaughton et~al.}(2016)\citenamefont{Connaughton,
  Burns, Goldstein, Blackburn, Briggs, Zhang, Camp, Christensen, Hui, Jenke
  et~al.}}]{connaughton2016fermi}
\bibinfo{author}{\bibfnamefont{V.}~\bibnamefont{Connaughton}},
  \bibinfo{author}{\bibfnamefont{E.}~\bibnamefont{Burns}},
  \bibinfo{author}{\bibfnamefont{A.}~\bibnamefont{Goldstein}},
  \bibinfo{author}{\bibfnamefont{L.}~\bibnamefont{Blackburn}},
  \bibinfo{author}{\bibfnamefont{M.}~\bibnamefont{Briggs}},
  \bibinfo{author}{\bibfnamefont{B.-B.} \bibnamefont{Zhang}},
  \bibinfo{author}{\bibfnamefont{J.}~\bibnamefont{Camp}},
  \bibinfo{author}{\bibfnamefont{N.}~\bibnamefont{Christensen}},
  \bibinfo{author}{\bibfnamefont{C.}~\bibnamefont{Hui}},
  \bibinfo{author}{\bibfnamefont{P.}~\bibnamefont{Jenke}},
  \bibnamefont{et~al.}, \bibinfo{journal}{The Astrophysical Journal Letters}
  \textbf{\bibinfo{volume}{826}}, \bibinfo{pages}{L6} (\bibinfo{year}{2016}).

\end{thebibliography}

\end{document}